\documentclass[fleqn,usenatbib]{mnras}

\usepackage{newtxtext,newtxmath}

\usepackage[T1]{fontenc}

\DeclareRobustCommand{\VAN}[3]{#2}
\let\VANthebibliography\thebibliography
\def\thebibliography{\DeclareRobustCommand{\VAN}[3]{##3}\VANthebibliography}

\usepackage{graphicx}	
\usepackage{amsmath}	
\usepackage{amssymb}	

\title[XUV-induced escape and habitability]{Stellar Flares versus Luminosity: XUV-induced Atmospheric Escape and Planetary Habitability}

\author[Atri et al.]{
Dimitra Atri,$^{1}$\thanks{E-mail: atri@nyu.edu}
Shane R. Carberry Mogan,$^{2,1}$
\\
$^{1}$Center for Space Science, New York University Abu Dhabi, PO Box 129188, Saadiyat Island, Abu Dhabi, UAE\\
$^{2}$Mechanical and Aerospace Engineering Department, New York University Tandon School of Engineering, 6 MetroTech Center, Brooklyn, NY 11201, USA
}

\date{Accepted XXX. Received YYY; in original form ZZZ}

\pubyear{2020}

\begin{document}
\label{firstpage}
\pagerange{\pageref{firstpage}--\pageref{lastpage}}
\maketitle

\begin{abstract}
Space weather plays an important role in the evolution of planetary atmospheres. Observations have shown that stellar flares emit energy in a wide energy range (10$^{30}$-10$^{38}$ ergs), a  fraction of which lies in X-rays and extreme ultraviolet (XUV). These flares heat the upper atmosphere of a planet, leading to increased escape rates, and can result in atmospheric erosion over a period of time. Observations also suggest that primordial terrestrial planets can accrete voluminous H/He envelopes. Stellar radiation can erode these protoatmospheres over time, and the extent of this erosion has implications for the planet's habitability. We use the energy-limited equation to calculate hydrodynamic escape rates from these protoatmospheres irradiated by XUV stellar flares and luminosity. We use the Flare-Frequency Distribution of 492 FGKM stars observed with {\it TESS} to estimate atmospheric loss in Habitable Zone planets. We find that for most stars, luminosity-induced escape is the main loss mechanism, with a minor contribution from flares. However, flares dominate the loss mechanism of $\sim$20\% M4-M10 stars. M0-M4 stars are most likely to completely erode both their proto- and secondary atmospheres, and M4-M10 are least likely to erode secondary atmospheres. We discuss the implications of these results on planetary habitability. 
\end{abstract}




\section{Introduction}
Planetary habitability is one of the most important concepts in exoplanet science. It is defined as the zone around a star in which a planet is able to sustain liquid water on its surface \citep{kasting1993habitable}. While this approach is useful to identify potentially hospitable planets around stars, it fails to take into account the damaging aspect of stellar activity on such planets. Stellar radiation in space weather events include stellar flares, Coronal Mass Ejections (CMEs), and Stellar Proton Events (SPEs), which emit energy in high-energy regimes of XUV (10-120 eV) photons and charged particles (10 MeV - 10 GeV) \citep{tylka2009new}. Space weather-induced effects include planetary atmospheric losses \citep{gronoff2020}, photochemistry \citep{tilley2019modeling}, as well as radiation enhancement on planetary surfaces \citep{2017MNRAS.465L..34A, 2020MNRAS.492L..28A}. A fraction of the stellar luminosity is emitted in the XUV range, which can similarly affect planetary atmospheres and surfaces. These effects can be of great importance for cool stars such as M-dwarfs whose habitable zones lie close to the host star ($\sim$0.01-1 AU) as this proximity makes potentially habitable planets highly vulnerable to space weather-induced damage. 

XUV radiation from a star heats up a planetary atmosphere and can lead to hydrodynamic escape. Other channels of escape include photochemical- and plasma-induced escape (e.g., \cite{gronoff2020} and references therein). However, studies have shown that thermal hydrodynamic escape is likely dominant in cases where a large amount of XUV energy is deposited (e.g., \citep{luger2015habitable}). Atmospheric escape can be estimated based on the total amount of heat deposited in the atmosphere, which is proportional to the XUV energy emitted from a star in steady-state or during a flare; e.g., via the energy limited escape formula \citep{watson1981dynamics}. Observations from {\it Kepler}, {\it Gaia}, and {\it TESS} have shown that active stars can emit energy in the 10$^{30}$-10$^{38}$ ergs energy range \citep{maehara2012superflares, notsu2019kepler, gunther2020}, a fraction of which is emitted in XUV. According to the Flare-frequency distribution (FFD) obtained from observations, the higher-energy flares, which can cause more damage, are rare, and the less effective lower-energy flares are more frequent. On long timescales, the atmospheric loss from flares will cumulatively lead to the erosion of the atmosphere, which can be estimated by using FFD obtained from observations.
 
\textit{Kepler} observations have shown that a large fraction of low-mass terrestrial planets have large H/He envelopes \citep{owen2016habitability}. It is believed that they accrete large H/He envelopes in early stages of their lives which can be eroded with time due to stellar radiation. \cite{wolfgang2015rocky} estimated this primordial atmospheric mass as $\sim$ 1\% of the mass of the planet. If these protoatmospheres can be completely stripped away, they can be replaced by secondary atmospheres, like we see in Solar System terrestrial planets (e.g., \cite{kopparapu2013, kopparapu2014}). Calculating the rate of atmospheric escape is therefore crucial in determining atmospheric structure and composition, which in turn has implications on planetary habitability. We focus our effort on recent data from {\it TESS} \citep{gunther2020}, which has given us the FFD of 492 FGKM stars. The main goal of this paper is to understand how stellar luminosity and flares can lead to atmospheric escape on habitable zone planets on long timescales and how these losses impact planetary habitability. 

\section{Method}
We obtain FFD from a sample of 492 FGKM stars observed with {\it TESS} \citep{gunther2020}, consisting of 128 M4-M10 stars, 233 M0-M4 stars, and 131 FGK stars, each of which has their own $\alpha$ and $\beta$ values derived from observations. We use these values to extrapolate FFD between 10$^{30}$ to 10$^{38}$ ergs:
\begin{equation}
    log_{10}(R_{F}) = \alpha log_{10}(E_\mathrm{bol}) + \beta.
\end{equation}
Here E$_\mathrm{bol}$ is the total bolometric energy of the flare and $R_{F}$ is the flare rate, which can also be expressed as:
\begin{equation}
    R_{F} = 10^{\beta} E_\mathrm{bol}^{\alpha}.
\end{equation} \label{eq:FFD}
We use this expression to calculate the occurrence rate of flares in this energy range over timescales of interest.

Next we use the catalog data to calculate star-planet distance by determining their HZs using the following method. All 492 of these stars have an effective temperature ($T_\mathrm{eff}$) derived in the data; however, only 485 stars have a derived effective radius ($R_\mathrm{eff}$). For those 7 stars with only $T_\mathrm{eff}$ available, as done by \cite{gunther2020}, $R_\mathrm{eff}$ is interpolated from the values given in \cite{pecaut2013}. A star's bolometric luminosity ($L_\mathrm{bol}$) is then calculated according to its $T_\mathrm{eff}$ and $R_\mathrm{eff}$:
\begin{equation}
L_\mathrm{bol} = 4 \pi R_\mathrm{eff}^2 \sigma T_\mathrm{eff}^4,    
\end{equation}
where $\sigma$ is the Stefan-Boltzmann constant. The star's $T_\mathrm{eff}$, $R_\mathrm{eff}$, and $L_\mathrm{bol}$ are then be used to determine its 6 habitable zones (HZ) limits \citep{kopparapu2013, kopparapu2014}: (1) Recent Venus, (2) Early Mars, (3) Runaway Greenhouse, and (4) Maximum Greenhouse for 1 $M_\oplus$; and Runaway Greenhouse for (5) 5 $M_\oplus$ and (6) 0.1 $M_\oplus$. Finally, the HZs in distance from the host star ($d_\mathrm{HZ}$) are calculated as:
\begin{equation}
d_\mathrm{HZ} = \sqrt{ \frac{L_\mathrm{bol} / L_\odot}{S_\mathrm{eff}}} \ [\mathrm{AU}],
\end{equation}
where $L_\odot$ is the stellar luminosity and $S_\mathrm{eff}$ is the effective stellar flux in the star's HZ.

From a star's bolometric energy, we can calculate the amount of XUV energy incident on top of a planetary atmosphere ($E_\mathrm{XUV}$) at the 6 $d_\mathrm{HZ}$:
\begin{equation}
E_\mathrm{XUV} = \frac{f_\mathrm{XUV} E_\mathrm{bol}}{4 \pi d_\mathrm{HZ}^{2}},
\end{equation}
where $f_\mathrm{XUV}$ is the estimated fraction of the total bolometric energy emitted in XUV. Since $f_\mathrm{XUV}$ depends on the spectral hardness, which can be $\sim$0.2 for hard-spectrum events \citep{woods2004solar}, and lower for soft-spectrum ones, we set it to 0.1 for all our events. After applying the star's FFD from Eq. \ref{eq:FFD} and integrating over the entire energy range of $10^{30}$-$10^{38}$ ergs, we calculate the mass losses at the 6 HZs via the energy-limited formulaton:
\begin{equation}
M_\mathrm{loss, flare} =  \sum_{10^{30}}^{10^{38}} \frac{\eta \pi R_{XUV}^2 E_\mathrm{XUV} R_{F}}{GM_{P} / R_{P}}.	
\end{equation}
Here $\eta$ is the heating efficiency of the atmosphere; $G$ is the gravitational constant; M$_{P}$ is mass of the planet; R$_{P}$ and $R_\mathrm{XUV}$ are the radius of the planet and the atmosphere's effective radius, respectively; and $\pi R_\mathrm{XUV}^2$ is the planetary envelope cross-section onto which the incident energy is absorbed. We further simplify this calculation by assuming $R_\mathrm{XUV} \sim R_{P}$:
\begin{equation}
M_{\mathrm{loss, flare}} \sim \sum_{10^{30}}^{10^{38}} \frac{\eta \pi R_{XUV}^3 E_\mathrm{XUV} R_{F}}{GM_{P}} = \sum_{10^{30}}^{10^{38}} \frac{\eta f_\mathrm{XUV} R_{P}^{3}}{4 d_\mathrm{HZ}^2 GM_{P}} 10^{\beta} E_\mathrm{bol}^{\alpha + 1}.
\end{equation}
\label{eq:M_loss_flare_energy_lim}
A range of values of heating efficiencies have been applied in literature, from as high as 100\% to as low as 10\% (\cite{shematovich2014heating} and references therein). We use a value of 10\% for our calculations based on the results of detailed modeling conducted by \cite{shematovich2014heating}, where it was concluded that values between 10-15\% would lead to accurate results. However, since we assume $R_\mathrm{XUV} \sim R_{p}$, even with a low heating efficiency our mass loss results may still be an underestimation. We apply Eq. \ref{eq:M_esc_lum_energy_lim} to 4 terrestrial planet sizes according to their HZs. That is, at HZs (1-4), $M_P = M_\oplus$ and $R_P = R_\oplus$, where $M_\oplus$ and $R_\oplus$ m are the mass and radius of Earth, respectively; and at HZ (5) and (6) we consider two ``super-Earths'' and a ``sub-Earth'' from the literature, respectively: $M_P = 5 M_\oplus$ and $R_P = 1.71 R_\oplus$ \citep{lammer2014origin}, $M_P = 5 M_\oplus$, and $R_P = 2.71 R_\oplus$ \citep{erkaev2016}, and $M_P = 0.1 M_\oplus$ and $R_P = 0.46 R_\oplus$ \citep{lammer2014origin}. We calculate the total mass losses over timescales of interest ($t$) at the 6 HZs as:
\begin{equation}
M_\mathrm{loss, flare} =  \sum  \left(\sum_{10^{30}}^{10^{38}} \frac{\eta f_\mathrm{XUV} R_{P}^{3}}{4 d_\mathrm{HZ}^2 GM_{P}} 10^{\beta} E_\mathrm{bol}^{\alpha + 1} \right) t.	
\end{equation}

The time-dependent XUV emission from a star can be obtained by using the following expressions \citep{luger2015habitable}:
    \[
    \frac{L_\mathrm{XUV}}{L_\mathrm{bol}} =
    \begin{cases}
    \left( \frac{L_\mathrm{XUV}}{L_\mathrm{bol}} \right)_\mathrm{sat}, & t \leq t_\mathrm{sat} \\
    \left( \frac{L_\mathrm{XUV}}{L_\mathrm{bol}} \right)_\mathrm{sat} \left( \frac{t}{t_\mathrm{sat}} \right)^{-\gamma}, & t > t_\mathrm{sat}
    \end{cases}
    \]
Here $\gamma=1.23$ is obtained from \cite{luger2015habitable} and applied to all star types and $t_\mathrm{sat}$ is the saturation time, which for FGK stars is 100 Myr \citep{luger2015habitable} and for M-dwarfs is 600 Myr \citep{gronoff2020}. $\frac{L_\mathrm{XUV}}{L_\mathrm{bol}}$ is difficult to calculate and is therefore set to $10^{-4}$ for M4-M10 stars and $10^{-3}$ for M0-M4 and FGK stars, based on recent estimates \citep{luger2015habitable, schmidt2015boss, wheatley2017strong}. We then use $L_\mathrm{XUV}$ to calculate the time-dependent stellar fluxes incident on the tops of the planets' atmospheres residing at the 6 HZs as:
\begin{equation}
    F_\mathrm{XUV} = \frac{L_\mathrm{XUV}}{4 \pi d_\mathrm{HZ}^2},
\end{equation}
which are then used to calculate $L_\mathrm{XUV}$-induced atmospheric escape rates ($\dot{M}_\mathrm{esc, L_\mathrm{XUV}}$) via the energy-limited formulaton:
\begin{equation}
  \dot{M}_{\mathrm{esc}, L_\mathrm{XUV}} \sim \frac{\eta \pi R_{XUV}^3 F_\mathrm{XUV}}{G M_{p}}.
\end{equation}
\label{eq:M_esc_lum_energy_lim}
Finally, total mass losses from these atmospheres are calculated by summing Eq. \ref{eq:M_esc_lum_energy_lim} over timescales of interest:
\begin{equation}
    M_{\mathrm{loss}, L_\mathrm{XUV}} = \sum  \dot{M}_{\mathrm{esc}, L_\mathrm{XUV}} t.
\end{equation}

\section{Results}
\begin{figure}
	\includegraphics[width=\columnwidth]{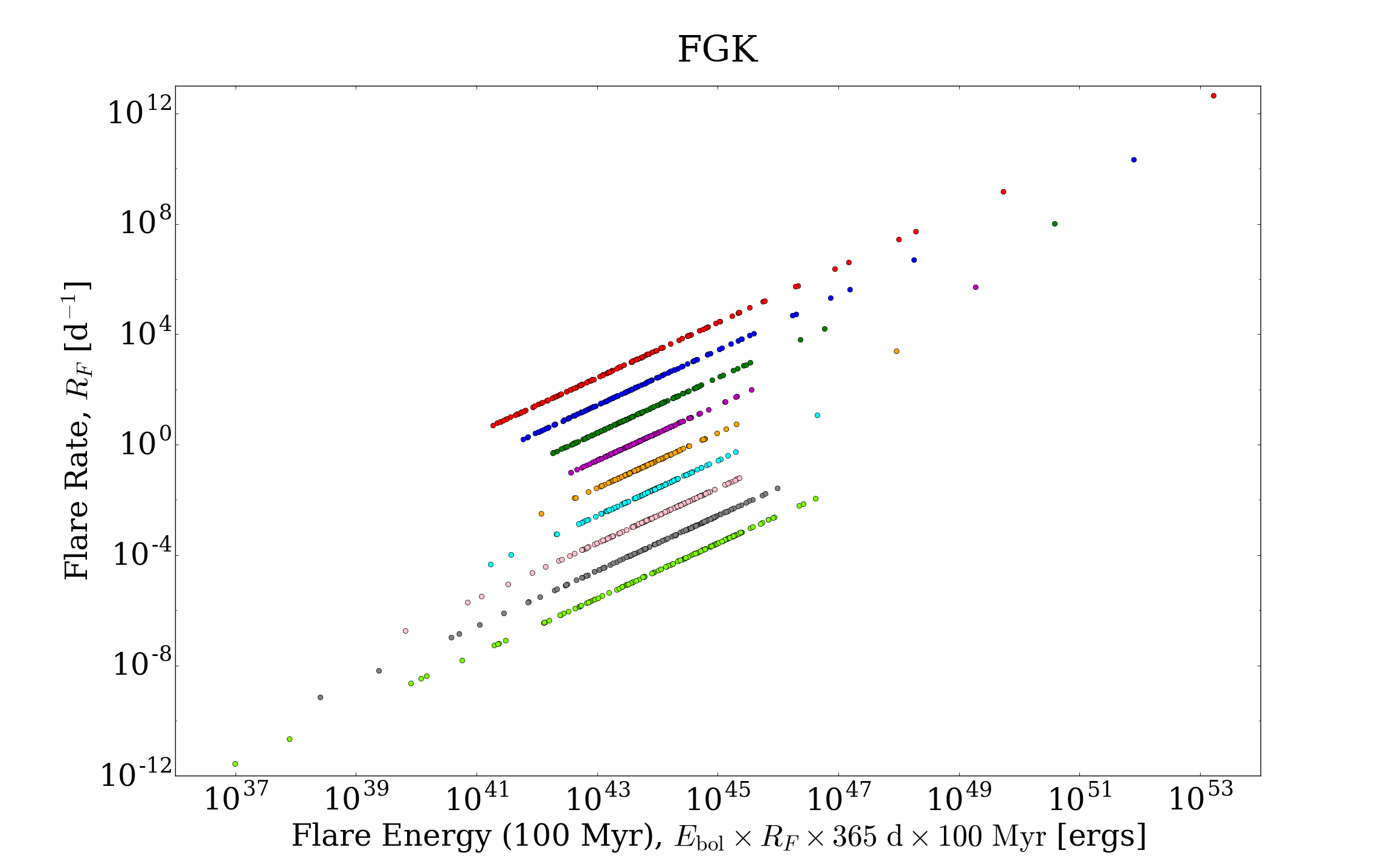}
    \caption{Total energy released by flares over 100 Myr corresponding to the flare's bolometric energy and frequency for FGK stars. Here the bolometric energy range is represented in red ($10^{30}$), blue ($10^{31}$), green ($10^{32}$), magenta ($10^{33}$), orange ($10^{34}$), cyan ($10^{35}$), pink ($10^{36}$), grey ($10^{37}$), lawn green ($10^{38}$).}
    \label{fig:FlareEnery_vs_FlareRate_FGK_100Myr}
\end{figure}

There is a large variation in the cumulative energy released from flares among the sample of 492 stars. Our calculations show that the total energy released by flares according to the flare energy ($10^{30}$-$10^{38}$) and corresponding daily rate ($R_F$) over 100 Myr can span over several orders of magnitude: for M4-M10 stars, $\sim$10$^{33}$-$10^{53}$ ergs; for M0-M4 stars, $\sim$10$^{28}$-$10^{50}$ ergs; and for FGK stars, $\sim$10$^{37}$-$10^{53}$ ergs (see Figure \ref{fig:FlareEnery_vs_FlareRate_FGK_100Myr}). If the drop off of $R_F$ from low to high energy flares is larger than the difference between their corresponding bolometric energies, then the contribution of these higher energy flares is minor on long timescales relative to the more frequent, lower energy flares. However, in some extreme cases, the high energy flares were frequent enough to significantly contribute to atmospheric losses.

\begin{figure}
	\includegraphics[width=\columnwidth]{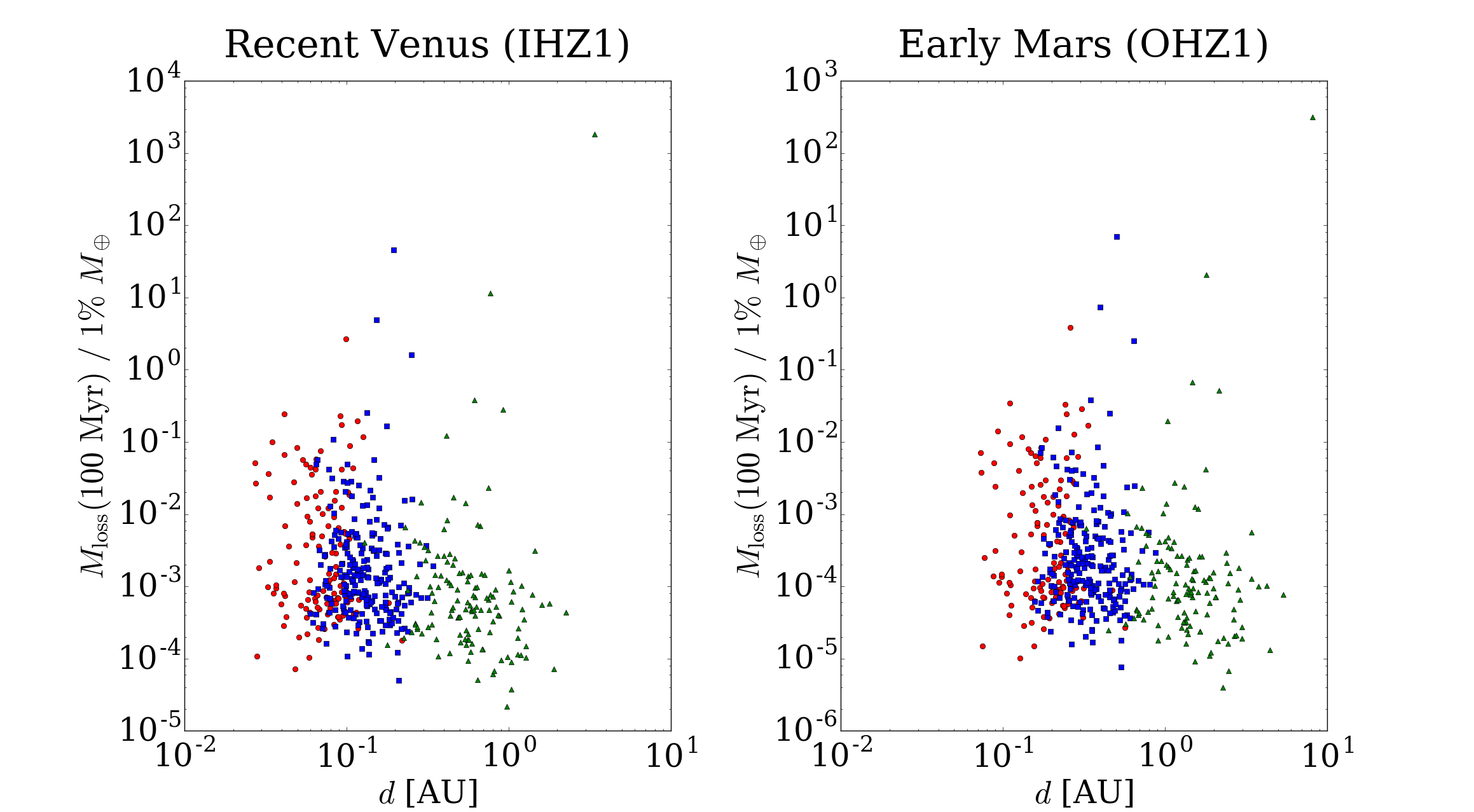}
    \caption{Flare-induced losses at Recent Venus (IHZ1) and Early Mars (OHZ1) HZs over 100 Myr relative to a $1 \% \ M_\oplus$ protoatmosphere. Red circles, blue squares, and green triangles represent M4-M10, M0-M4, and FGK stars, respectively.}
    \label{fig:Flare_Loss_100Myr}
\end{figure}

\begin{figure}
	\includegraphics[width=\columnwidth]{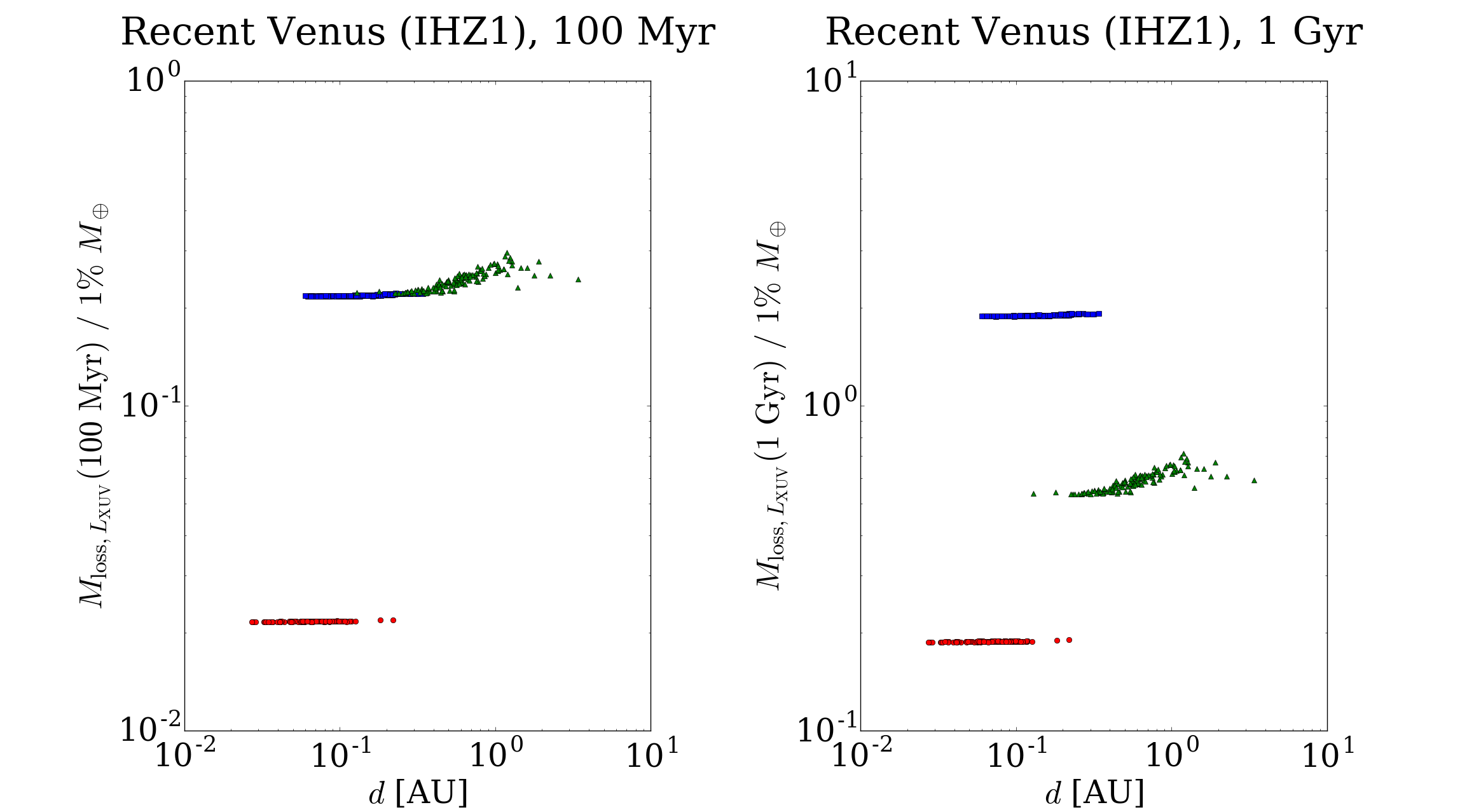}
    \caption{$L_\mathrm{XUV}$-induced losses at Recent Venus (IHZ1) over 100 Myr (\textit{left}) and 1 Gyr (\textit{right}) relative to a $1 \% \ M_\oplus$ protoatmosphere. Red circles, blue squares, and green triangles represent M4-M10, M0-M4, and FGK stars, respectively. Here the upper bound of $L_\mathrm{XUV}$ is applied for M4-M10 and M0-M4 stars: $L_\mathrm{XUV}/L_\mathrm{bol} = 10^{-4}$ and $L_\mathrm{XUV}/L_\mathrm{bol} = 10^{-3}$, respectively.}
    \label{fig:Lum_Loss_100Myr}
\end{figure}

Next, we show total mass losses over 100 Myr and 1 Gyr time periods with the three star categories. Mass loss is expressed in terms of the mass of the planet's protoatmosphere, which as we described earlier is 1\% of the core mass. 
Figure \ref{fig:Flare_Loss_100Myr} shows flare-induced atmospheric loss for planets in IHZ1 (Recent Venus) and OHZ1 (Early Mars), over 100 Myr. It can be seen that 6 stars (1 M4-M10, 3 M0-M4, 2 FGK) are able to completely deplete the primordial atmosphere in 100 Myr in IHZ1, and only 3 stars (1 M0-M4, 2 FGK) in OHZ1. Figure \ref{fig:Lum_Loss_100Myr} shows XUV stellar luminosity-induced atmospheric loss over 100 Myr and 1 Gyr in IHZ1, where the former timescale is before the saturation phase, which is 100 Myr for FGK stars and 600 Myr for M-dwarfs. Although FGK stars induce the most losses over 100 Myr, followed closely by M0-M4, none of them are able to completely deplete the atmosphere. Over 1 Gyr, however, all M0-M4 stars are able to completely deplete the atmosphere. The comparative loss of FGK stars is significantly lower because M0-M4 stars have high irradiation levels in XUV ($L_\mathrm{XUV}/L_\mathrm{bol} = 10^{-3}$) coupled with a longer saturation time.

\begin{figure}
	\includegraphics[width=\columnwidth]{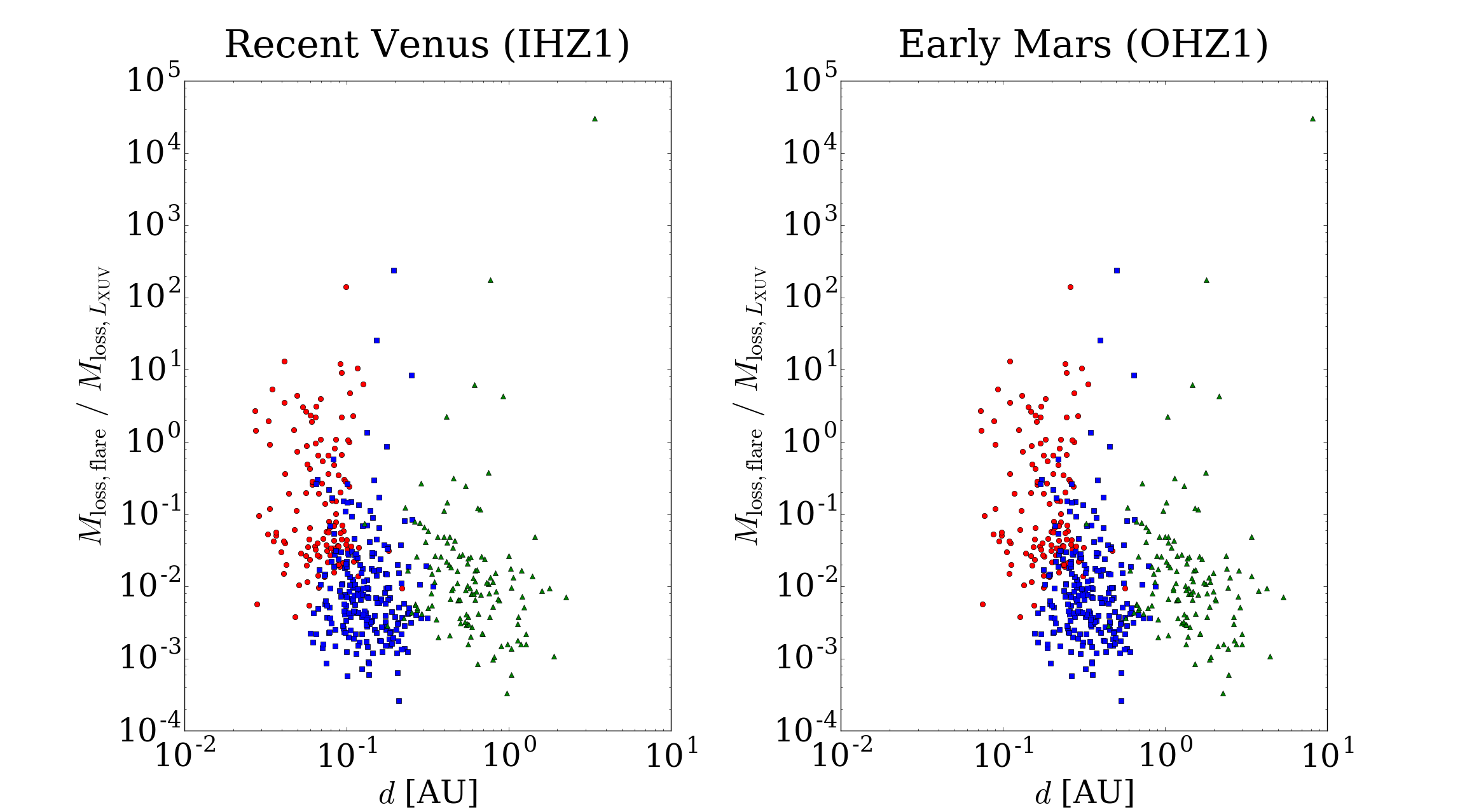}
    \caption{Ratios of Flare- to $L_\mathrm{XUV}$-induced losses at Recent Venus (IHZ1) and Early Mars (OHZ1) HZs over 1 Gyr. Red circles, blue squares, and green triangles represent M4-M10, M0-M4, and FGK stars, respectively.}
    \label{fig:FlareOverLum_Loss_1Gyr}
\end{figure}

\begin{figure}
	\includegraphics[width=\columnwidth]{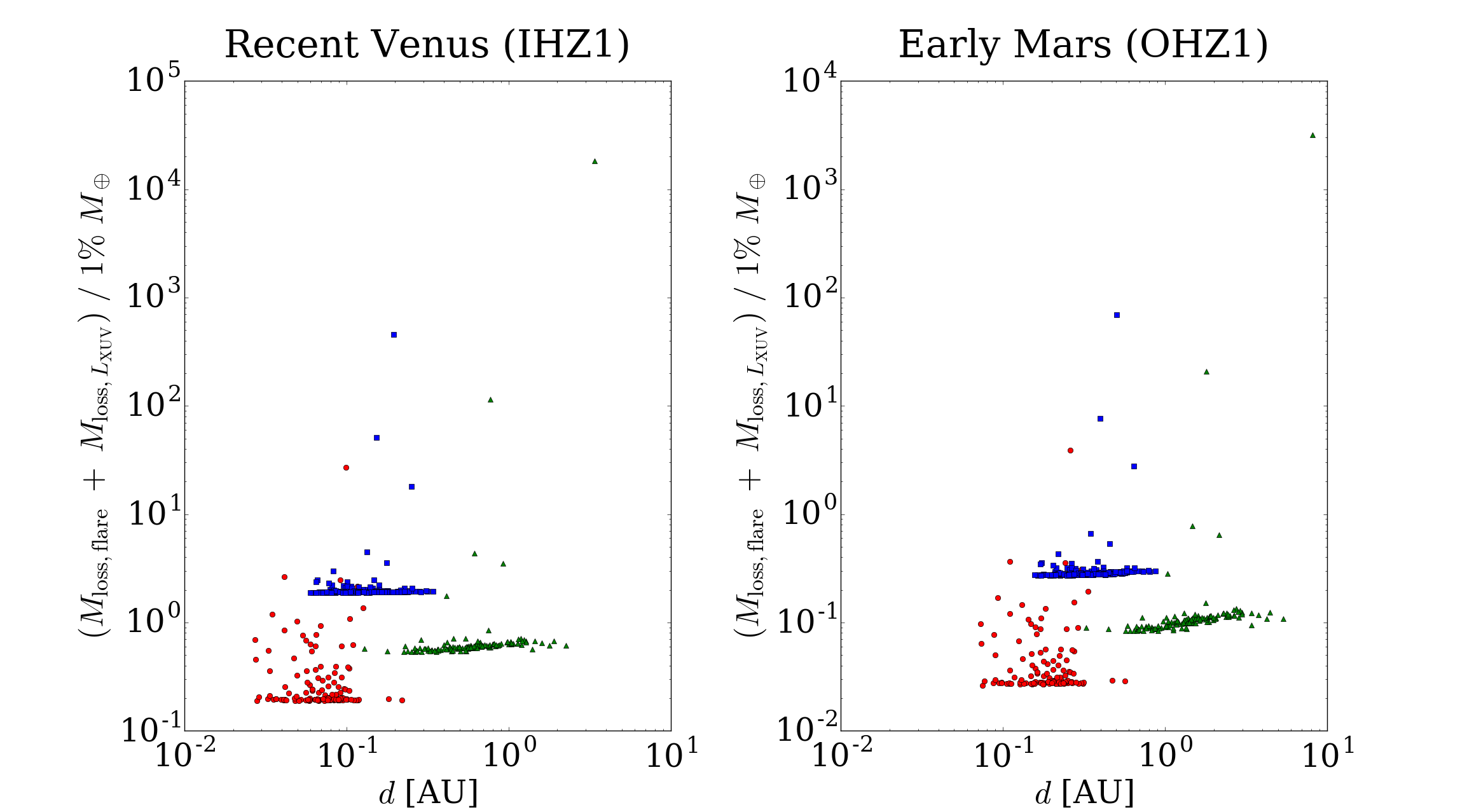}
    \caption{Sum of Flare- and $L_\mathrm{XUV}$-induced losses at Recent Venus (IHZ1) and Early Mars (OHZ1) HZ over 1 Gyr relative to a $1 \% \ M_\oplus$ protoatmosphere. Red circles, blue squares, and green triangles represent M4-M10, M0-M4, and FGK stars, respectively.}
    \label{fig:FlarePlusLum_Loss_1Gyr}
\end{figure}

\begin{figure}
    \includegraphics[width=\columnwidth]{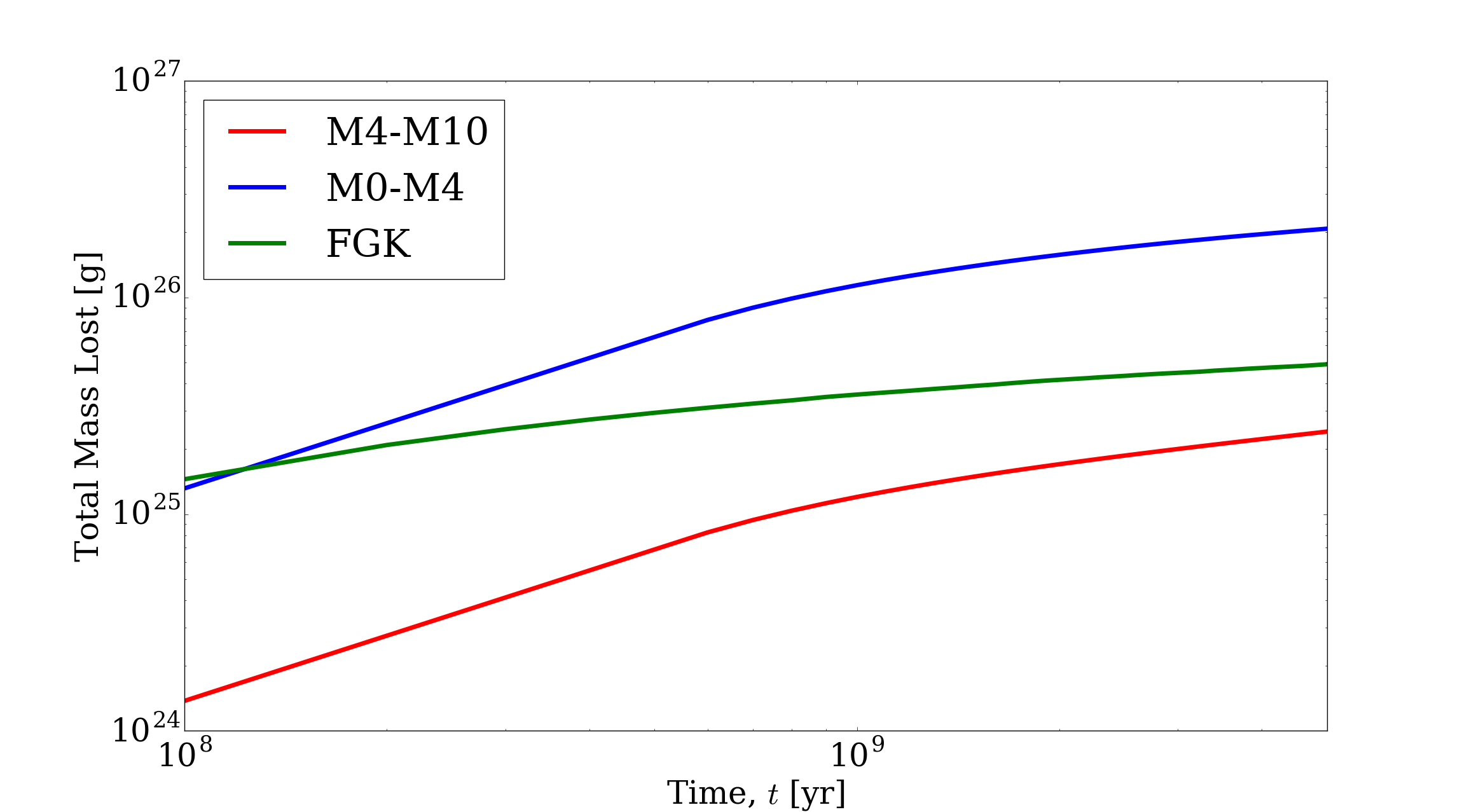}
    \caption{Median sums of Flare- and $L_\mathrm{XUV}$-induced losses at Recent Venus (IHZ1) over 5 Gyr, where red, blue, and green solid lines represent M4-M10, M0-M4, and FGK stars, respectively.}
    \label{fig:MedianMassLost_5Gyr}
\end{figure}

We now compare atmospheric loss from flares and stellar luminosity. Figure \ref{fig:FlareOverLum_Loss_1Gyr} shows the ratio of flare- to luminosity-induced losses over 1 Gyr in IHZ1 and OHZ1. In both cases, it can be seen that flaring is the dominant source of escape in a significant fraction of M4-M10 stars, and to a lesser degree in other categories over 1 Gyr: $\sim$20\% (26/128), $\sim$2\% (4/233), and $\sim$4\% (5/129) of the M4-M10, M0-M4, and FGK stars, respectively. In Figure \ref{fig:FlarePlusLum_Loss_1Gyr}, we show the total atmospheric loss from flares and luminosity over 1 Gyr in IHZ1 and OHZ1. While all M0-M4 stars and a small number in the other categories were able to completely erode the primordial atmosphere in IHZ1, a majority of all of the stars in OHZ1 were not. We summarize these results in Table \ref{table:loss} where the median mass loss is given for the three star categories in 6 HZs. It can be seen that the median luminosity-induced loss is about an order of magnitude higher than flare-induced loss for M4-M10, and about two orders of magnitude higher for M0-M4 and FGK stars respectively. Finally, in Figure \ref{fig:MedianMassLost_5Gyr}, we show the time evolution of the median mass loss from the three categories of stars over 5 Gyr. The rise in losses from FGK stars is considerably lower than that of M-dwarfs because their saturation times are much smaller and hence the XUV decline starts much earlier as described previously.    

\begin{table}
\centering
\caption{Median losses for each star type at all 6 HZs over 1 Gyr. For IHZ3, losses from the the more dense super-Earth ($M_P = 5 M_\oplus$, $R_P = 2.71 R_\oplus$) are shown.}
 \begin{tabular}{lcc}
  \hline
  M4-M10 & Flares & $L_\mathrm{XUV}$ \\
  & ($\times 10^{23}$) [g] & ($\times 10^{24}$) [g] \\
  \hline
  IHZ1 & 7.57 & 11.2 \\
  OHZ1 & 1.08 & 1.61 \\
  IHZ2 & 4.72 & 7.01 \\
  OHZ2 & 1.21 & 1.79 \\
  IHZ3 & 20.2 & 30.0 \\
  IHZ4 & 4.08 & 6.07 \\
  \hline
  M0-M4 & Flares & $L_\mathrm{XUV}$ \\
  & ($\times 10^{23}$) [g] & ($\times 10^{25}$) [g]\\
  \hline
  IHZ1 & 7.49 & 11.3\\
  OHZ1 & 1.10 & 1.65 \\
  IHZ2 & 4.67 & 7.04 \\
  OHZ2 & 1.22 & 1.84 \\
  IHZ3 & 20.0 & 30.1 \\
  IHZ4 & 4.04 & 6.10 \\  
  \hline
  FGK & Flares & $L_\mathrm{XUV}$\\
  & ($\times 10^{23}$) [g] & ($\times 10^{25}$) [g] \\
  \hline
  IHZ1 & 3.28 & 3.46 \\
  OHZ1 & 0.564 & 0.597 \\
  IHZ2 & 2.05 & 2.16 \\
  OHZ2 & 0.628 & 0.664 \\
  IHZ3 & 8.74 & 9.22 \\
  IHZ4 & 1.78 & 1.88 \\  
  \hline  
 \end{tabular}
\end{table} \label{table:loss}

\section{Conclusion and Discussion}
The ability of a planet to retain its atmosphere is one of the main factors governing its habitability. We have investigated XUV-induced atmosopheric loss from flares and luminosity on different star types in 6 cases of HZs. We applied the energy-limited escape formula to calculate hydrodynamic escape. We found that more frequent, lower energy flares tended to dominate as a source of mass loss over long timescales, whereas flares with energies beyond $10^{36}$ ergs (superflares) do not make a significant contribution because of their low occurance rate, except in a limited number of cases (Figure \ref{fig:FlareEnery_vs_FlareRate_FGK_100Myr}). We only show calculations for total losses induced from these flares over $10^8 - 10^9$ years. However, if we were concerned with calculating instantaneous losses from an individual flare then we would factor in its duration, $t_F$, according to its bolometric energy via the following equation (\citep{tilley2019modeling}):
\begin{equation}
    t_F = 10^{0.395 \log_{10} (E_\mathrm{bol} - 9.269)},
\end{equation}
which gives durations of $\sim 4 \times 10^2 - 5 \times 10^5$ s for $E_\mathrm{bol} = 10^{30} - 10^{38}$ ergs. These flare durations can exceed by more than an order of magnitude that of a recent study for Hot Jupiters \citep{bisikalo2018atmospheric}. Indeed, these larger flare durations and the much smaller planetary radii in this study compared to that of the Hot Jupiter study make it difficult to draw direct comparisons.

We have demonstrated that for most stars, luminosity-induced escape is the main loss mechanism, with only a minor contribution from flares. However, flares dominate the loss mechanism of $\sim$20\% M4-M10 stars. M0-M4 stars, because of their high XUV irradiation levels, are able to completely erode the planetary protoatmospheres in all inner HZs, including IHZ3 where the core mass is 5 $M_\oplus$. This contrasts with previous studies (e.g., \cite{erkaev2016} and references therein) that suggested protoplanetary cores with masses $\geq M_\oplus$ orbiting inside the HZs of M-type stars are likely to keep their hydrogen envelopes. However, when calculating $L_\mathrm{XUV}$-induced losses, we assumed escape was always hydrodynamic, which \cite{owen2016habitability} demonstrated can lead to significant overestimations. If indeed our $L_\mathrm{XUV}$-induced losses are overestimations, then our result that flares can induce greater losses is even more significant and likely occurs for more stars. Also the lack of atmospheric erosion for protoplanetary cores with masses in the range of 0.1-5 $M_\oplus$ orbiting FGK (solar-like) stars within their HZs agrees with \cite{erkaev2016}. 

M0-M4 stars are the most likely ones to erode secondary atmospheres, as seen in Figure \ref{fig:FlarePlusLum_Loss_1Gyr}, because they are able to maintain relatively high XUV irradiation levels over long timescales. A combination of high XUV from stellar luminosity and flares and small star-planet distance makes these planets especially sensitive to erosion of both proto- and secondary atmospheres. On the other hand, the least likely ones to erode their secondary atmospheres are M4-M10 stars, which consisted of the most stars whose primary loss mechanism was flare-induced. These results have significant implications for planetary habitability because about 75\% of stars in the Milky Way are M-dwarfs \citep{owen2016habitability} and observations suggest that they host twice the number of planets around them compared to other stars \citep{hardegree2019kepler}. Moreover, the extended atmospheres of these planets are more capable of being observed due to the relatively small ratio of planetary to stellar radius \cite{lammer2011uv}. Therefore, based on the substantial losses that can be induced by XUV flares and luminosity at the close-in HZs of the M-type stars, addressing the corresponding expansion of these atmospheres could be useful for observing potentially habitable planets around M dwarfs. However, we did not address this and simply assumed that $R_\mathrm{XUV} \sim R_P$ in our energy-limited escape calculations. Also, most planets in OHZ1 ($\sim$ 99\%) are not significantly impacted by stellar radiation-induced atmospheric loss. This also signifies the importance of planetary composition and structure, which governs the location of the HZ.

There were several caveats in our formulation. The energy-limited equation has been shown to work particularly well for hydrodynamic escape driven by the stellar XUV flux but has been shown to underestimate mass loss for highly irradiated, low-density planets, where escape is driven by a combination of the planetary intrinsic thermal energy and low gravity; and overestimate mass loss for planets with hydrostatic atmospheres in which it is controlled by Jeans escape (\cite{kubyshkina2018} and references therein). Given the large amount of XUV energy from flares deposited into the protoatmospheres of the terrestrial planets at their close proximities to the host star, we expect the ensuing escape to be hydrodynamic. We treat two flares totally independent of each other, which might not be the case for low-energy flares with high occurance rates. As previously mentioned, we did not apply a more complex absorption/upper atmospheric model to determine $R_\mathrm{XUV}$. We applied a constant heating efficiency up to 5 Gyr despite it being shown to vary with time (e.g., \cite{murray2009atmospheric}). Moreover, we did not consider any possible radiative cooling that could affect the efficiency (e.g., \cite{owen2016habitability}). The energy-limited escape formula does not account for neutral losses of an atmosphere, such as dissociation and ionization, as well as subsequent recombination. \cite{erkaev2016} simulated EUV-driven mass loss of protoatmospheres and took into account dissociation and ionization of molecular hydrogen and recombination of atomic hydrogen by solving more complex fluid equations for mass, momentum, and energy conservation. They found that energy-limited escape with $R_\mathrm{XUV} \sim R_P$ and $R_\mathrm{XUV} > R_P$ were lower and upper bounds, respectively, and within an order of magnitude of the sum of atomic and molecular hydrogen escape (see Figure 1 therein). Only about half of the 492 stars (10 M4-M10, 133 M0-M4, and 110 FGK) from the \textit{TESS} data had information from which we could derive stellar masses. We used these masses to calculate the Roche lobe radii and corresponding stellar tidal effects \citep{erkaev2007} and found that $K > 0.95$ for all of the stars. Therefore, due to this relatively small enhancement of escape as well as the fact that only about half the stars had sufficient data to make such a calculation, we neglected this variable when solving the energy-limited escape formula. We use $T_\mathrm{eff}$ and $R_\mathrm{eff}$ from \textit{TESS} data to calculate the stars' current luminosities to determine HZs. However, we do not consider how the luminosity and thus HZs could vary with time nor where in its lifetime these stars may be (e.g., \cite{luger2015habitable}). Our study focuses on erosion or protoatmospheres but we do not consider processes that could protect an atmosphere from being lost to space (e.g., \cite{johnstone2019}). Finally, our study is primarily concerned with the enhanced thermal escape XUV stellar flares can induce; however, we do not consider additional nonthermal escape channels, e.g., charge-exchange interactions between atmospheric neutrals and stellar wind protons.

Atmospheric escape is a complex process and we need a better numerical modeling approach to estimate the total loss by including other channels of escape. We need more flare observations to get a better picture of FFD around a variety of stars. More observations of escaping atmospheres will help in better contraining atmospheric models and aid in understanding the long-term effects of stellar activity on planetary atmospheres and implications on their habitability.   

\section*{Acknowledgements}

DA acknowledges support from the New York University Abu Dhabi (NYUAD) Institute research grant G1502. SRCM acknowledges support from the NYUAD Global PhD Fellowship. 



\section*{Data Availability}
The data underlying this article are available at: \url{https://iopscience.iop.org/1538-3881/159/2/60/suppdata/ajab5d3at1_mrt.txt}

\bibliographystyle{mnras}
\bibliography{example} 








\end{document}